\documentclass[12pt,a4paper,final]{iopart}

\renewcommand{\figurename}{Figure}
\expandafter\let\csname equation*\endcsname\relax

\expandafter\let\csname endequation*\endcsname\relax
\usepackage{graphicx}
\usepackage{dcolumn}
\usepackage{bm,soul}
\usepackage{color}
\usepackage{ amssymb }
\usepackage{subfigure}   
\usepackage{ amsmath }

\begin{document}
 
\title[Hysteretic uncertainty relation]{Hysteretic thermodynamic uncertainty relation for systems with broken time-reversal symmetry}

\author{Karel Proesmans}
 \ead{Karel.Proesmans@uhasselt.be}
 \address{Hasselt University, B-3590 Diepenbeek, Belgium.}
\author{Jordan M. Horowitz}
\address{Department of Biophysics, University of Michigan, Ann Arbor, Michigan, 48109, USA}
\address{Center for the Study of Complex Systems, University of Michigan, Ann Arbor, Michigan 48104, USA}

\date{\today}

\begin{abstract}
The thermodynamic uncertainty relation {bounds the amount current fluctuations can be suppressed in terms of the dissipation} in a mesoscopic system. 
By considering the fluctuations in the hysteresis of the current -- the sum of the currents in the time-forward and time-reversed processes -- we extend this relation to systems with broken time-reversal symmetry, either due to the presence of odd state variables, odd driving fields or due to explicit time-dependent driving that is time-reversal asymmetric.
We illustrate our predictions on a dilute, weakly-interacting gas driven out of equilibrium by the slow compression of a piston and on a ballistic multi-terminal conductor with an external magnetic field.
\end{abstract}


\section{Introduction}

In  fluctuating systems, stochastic thermodynamics has emerged as a systematic theoretical framework for identifying the thermodynamic properties of currents -- such as heat, particle transport and work~\cite{seifert2012stochastic,van2015ensemble}.
Of particular interest is identifying universal energetic trade-offs (or constraints) to such flows, as they often imply general performance bounds on mesoscopic devices~\cite{funo2015work,brandner2015thermodynamics,proesmans2015onsager,proesmans2016power,shiraishi2016universal}.
One such result is the thermodynamic uncertainty relation, which bounds the fluctuations in the accumulation of a generic current $J_T$ in time $T$ with the system's entropy production rate $\sigma$. 
In its original formulation it states that for a system with Markovian dynamics the relative fluctuations -- given as the ratio of the average steady-state flux  $\langle j\rangle =\lim_{T\to\infty}\langle J_T\rangle/T$ and variance $\textrm{Var}(j)=\lim_{T\to\infty}{\rm Var}(J_T)/T$ -- is bounded as~\cite{barato2015thermodynamic,gingrich2016dissipation},
\begin{equation}
    \frac{\left\langle j\right\rangle^2}{\textrm{Var}(j)}\leq \frac{\sigma}{2k_B},\label{tur1}
\end{equation}
where $k_B$ is Boltzmann's constant.
This prediction is based on two necessary assumptions:
(i) the system relaxes to a unique time-independent steady state, and (ii) the dynamics are { time-reversal symmetric}, \emph{i.e.}, the system's configurations and all the forces {are even and do not change sign  under time reversal, which forbids the presence of, say, magnetic fields}.
Under these conditions, several extensions have been formulated over the last few years, leading to thermodynamic uncertainty relations for finite-time processes \cite{pietzonka2017finite,horowitz2017proof,dechant2018current}, equilibrium systems \cite{guioth2016thermodynamic}, non-Markovian systems \cite{knoops2018motion}, and formulations in terms of other variables \cite{garrahan2017simple,gingrich2017fundamental,nardini2018process,di2018kinetic}. Furthermore, the above relation has found applications in the study of chemical reaction networks \cite{barato2015universal,pietzonka2016universal2,proesmans2018case}, molecular motors \cite{pietzonka2016universal,hwang2018energetic,wierenga2018quantifying,brown2018pulling}, heat engines \cite{pietzonka2018universal,solon2018phase,vroylandt2018degree}, self-assembly \cite{nguyen2016design}, synchronization \cite{PhysRevE.98.032119}, and inference of entropy production \cite{gingrich2017inferring,li2018quantifying}. It has also been related to information-theoretic measures \cite{ito2018stochastic,hasegawa2018information} and was an inspiration for a new measurement method for diffusion coefficients~\cite{dechant2018estimating}.

If we relax the steady-state assumption and allow time-dependent driving, it is known that the standard thermodynamic uncertainty relation Eq.~(\ref{tur1}) fails~\cite{barato2016cost,rotskoff2017mapping,proesmans2017discrete,ray2017dispersion,1367-2630-20-10-103023,chiuchiu2018mapping,barato2018unifying,koyuk2018generalization,dechant2018entropic,holubec2018cycling}. 
However, if one focuses on time-symmetric, periodic driving, say with period $\Delta t$, then one can  show a weaker bound in the periodic steady state \cite{proesmans2017discrete},
{
\begin{equation}
    \frac{\left\langle j_{\Delta t}\right\rangle^2}{{\rm Var}(j_{\Delta t})}\leq \frac{1}{2\Delta t}\left(\exp\left(\frac{\Delta S}{k_B}\right)-1\right)\label{tur2}
\end{equation}
}
where $\Delta S$ is the total entropy production over one period 
with $\langle j_{\Delta t}\rangle=\lim_{n\to\infty}\langle J_{n\Delta t}\rangle/n$ and $ \textrm{Var}\left(j_{\Delta t}\right)=\lim_{n\to\infty}{\rm Var} (J_{n\Delta t})/n$ the average current and variance per period obtained in the limit that the number of periods $n$ tends to infinity.
This relation also holds for discrete-time steady-state Markov chains, where $\Delta S$ is the average entropy production per step and $\Delta t$ its duration. 
Several alternatives have been proposed for the case where the driving is not time-reversal symmetric~\cite{1367-2630-20-10-103023,barato2018unifying,koyuk2018generalization}. These alternatives rely on the introduction of new quantities that mix thermodynamic with kinematic properties.
As a result, the interpretation of their physical meanings can be less transparent than entropy production. 

Similar problems arise for steady-state systems with broken time-reversal symmetry, either due to the presence of odd variables or odd external fields. 
This problem can already be seen in the linear response regime in the presence of a magnetic field, where Eq.~(\ref{tur1}) no longer holds due to the breakdown of Onsager symmetry \cite{PhysRevLett.121.130601}.
With odd variables, like the momentum in underdamped Brownian motion, the issues are more subtle as one usually encounters a reversible Hamiltonian evolution perturbed by thermal fluctuations.
Thus, currents can flow in phase space even in the absence of dissipation making the validity of Eq.~(\ref{tur1}) suspect.
The result is that the thermodynamic uncertainty relation cannot be adapted straightforwardly to underdamped dynamics \cite{maes2017frenetic,fischer2018large,van2019uncertainty}. Furthermore, quantum tests of the thermodynamic uncertainty relation that naturally combine Hamiltonian evolution with dissipative dynamics have violated Eq.~\eqref{tur1}~\cite{agarwalla2018assessing,PhysRevB.98.085425,carrega2019optimal} or required modification and weakening~\cite{brandner2018thermodynamic,Guarnieri2019thermodynamics}.

In this article, we develop a modified thermodynamic uncertainty relation that overcomes some of these difficulties by taking into account the fluctuations in the current of the time-reversed dynamics ${\tilde J}_T$.
In particular, we find that the fluctuations in the sum of the current and its time reversal or \emph{hysteretic} current $J_T+{\tilde J}_T$ satisfies a bound of the form
\begin{equation}\label{eq:newTUR}
\frac{\langle J_T+{\tilde J}_T\rangle^2}{{\rm Var}(J_T)+{\rm Var}({\tilde J}_T)}\le \exp\left(\frac{{\mathcal R}+\tilde{\mathcal R}}{2}\right)-1,
\end{equation}
where ${\mathcal R}$ and $\tilde{\mathcal R}$ are particular path averages of the stochastic process and its time-reverse, related to the entropy flow into the environment.
We further find particularly transparent interpretations upon specializing to isothermal driven processes or nonequilibrium steady-states within linear response.

\section{Setup}

Our analysis applies to mesoscopic systems that can be modeled as Markov processes.  In this section, we set the stage by reviewing the relevant properties of their dynamics and thermodynamics, before introducing our main theoretical tool, level 2 large deviations for independent trajectories.

\subsection{Dynamics and environmental entropy flow}

We have in mind a mesoscopic system with configurations labeled by the vector of state variables ${\bf x}=\{x_1,\dots,x_M\}$ whose evolution is driven by a collection of controllable external parameters or forces $\boldsymbol\lambda=\{\lambda_1,\dots,\lambda_\Theta\}$.
Both the state variables  and control parameters are further assumed to have a well defined symmetry under time-reversal, which we indicate as $\boldsymbol\epsilon{\bf x}=\{\epsilon_1 x_1,\dots, \epsilon_M x_M\}$  and $\boldsymbol\epsilon\boldsymbol\lambda=\{\epsilon_1 \lambda_1,\dots, \epsilon_\Theta \lambda_\Theta\}$ with $\epsilon_i=1$ for even variables/parameters, such as positions, and $\epsilon_i=-1$ for odd variables/parameters, such as velocities or magnetic fields.

Due to thermal fluctuations, the evolution of the system is stochastic.
Thus, in each realization of the process over a time interval $t\in[0,T]$ the system traces out a random trajectory $\Gamma=\{{\bf x}(t)\}_{t=0}^T$, which we allow to also depend on a specified driving protocol for the parameters $\boldsymbol\lambda(t)$.
We concisely denote the probability for this trajectory as
\begin{equation}
p_\Gamma=P[\Gamma|{\bf x}(0);\boldsymbol\lambda(t)]\rho_0[{\bf x}(0)]
\end{equation}
in terms of an initial probability density $\rho_0$ and the conditional probability  $P$ to observe the trajectory under the driving protocol given the initial configuration.

Conjugate to this forward process is a \emph{reverse} process characterized by a time-reversed driving protocol $\tilde{\boldsymbol\lambda}(t)=\boldsymbol\epsilon\boldsymbol\lambda(T-t)$.
Allowing for arbitrary initial probability distribution $\tilde\rho_0$, we have for the reverse trajectory probability
\begin{equation}
{\tilde p}_\Gamma=P[\Gamma|{\bf x}(0);\tilde{\boldsymbol\lambda}(t)]\tilde\rho_0[{\bf x}(0)],
\end{equation}

where the tilde on ${\tilde p}_\Gamma$ emphasizes that this is a distinct probability on all trajectories $\Gamma$ characterized by the time-reversed driving $\tilde{\boldsymbol\lambda}(t)$.

Having introduced the trajectory probability for the forward process and the reverse process, we can now lean on the theoretical framework of stochastic thermodynamics~\cite{seifert2012stochastic} to connect these trajectory dynamics to  thermodynamics.
We assume that our dynamics satisfy the principle of local detailed balance, which allows us to identify the entropy flow into the environment with the time-reversal asymmetry of the trajectory probabilities.
In particular, the entropy flow can be identified by comparing the probability of a trajectory $\Gamma$ in the original forward process to the time-reversed trajectory $\tilde\Gamma=\{\boldsymbol\epsilon {\bf x}(T-t)\}_{t=0}^T$ in the reverse process with time-reversed driving \cite{seifert2012stochastic}:
\begin{equation}\label{eq:Ent}
\Delta S_{\rm env}=k_{\rm B}\left\langle\ln\frac{P[\Gamma|{\bf x}(0);\boldsymbol\lambda(t)]}{P[\tilde\Gamma|\boldsymbol\epsilon{\bf x}(T);\tilde{\boldsymbol\lambda}(t)]}\right\rangle_{{\boldsymbol\lambda}(t)},
\end{equation}
where the average is over the forward driving.
Once we specify the thermodynamic reservoirs in the environment, the entropy flow can be related to the flow of energy through system.
For example, if the environment is composed of a single thermal reservoir at inverse temperature $\beta$, then the entropy flow is simply proportional to the heat $Q$ exhausted by the system: $\Delta S_{\rm env}=\beta Q$.

The identification of the environmental entropy flow in the presence of odd variables from the time-reversal asymmetry of the path probability as in Eq.~\eqref{eq:Ent} requires some comment.  
Microscopic reversibility is not sufficient to make the identification in Eq.~\eqref{eq:Ent} when there are odd-parity variables~\cite{spinney2012nonequilibrium}.
However, it has been pointed out that if the environment is composed of thermodynamic reservoirs, then additional symmetries, such as rotational symmetry or spatial parity, allow one to correctly identify the environmental entropy change through the time-reversal asymmetry of the trajectory probability \cite{deffner2015jarzynski,van2015stochastic,horowitz2016work}.
Throughout this paper we will assume such an identification to be possible.

\subsection{Level 2 large deviations for independent trajectories}

Our main approach to obtaining the inequality in Eq.~\eqref{eq:newTUR} will be to analyze the large deviations in the trajectory fluctuations.
In this section, we provide a heuristic argument underpinning this approach and review relevant formulas. One could cast the following argument into a more rigorous formulation via a limiting procedure, as was done in \cite{barato2018unifying}.

Let us imagine that we take $N\gg 1$ copies of our system, each initialized in $\rho_0$ and let to evolve independently for a time $T$ under the same driving protocol $\boldsymbol\lambda(t)$.
Each copy will trace out a different trajectory $\Gamma_i$, with $i=1,\dots,N$.
We can ask what fraction of the copies trace out the same trajectory $q_\Gamma=(1/N)\sum_i \delta(\Gamma-\Gamma_i)$, thereby forming an empirical (or observved) trajectory density.
Since each copy is independent, we have, roughly, for the probability distribution of the empirical density
\begin{equation}
{\mathcal P}_N(\{q_\Gamma\})=\frac{N!}{\prod_{\Gamma}(Nq_\Gamma)!}\left(p_{\Gamma}\right)^{Nq_\Gamma}\sim e^{-N{\mathcal J}(\{q_\gamma\})},
\end{equation}
where we have identified the large deviation rate function from the exponential decay of rare fluctuations \cite{touchette2009large}
\begin{equation}
\mathcal{J}(\{q_{\Gamma}\})=-\lim_{N\to\infty}\frac{1}{N}\ln  {\mathcal P}_N(\{q_{\Gamma}\})=\sum_\Gamma q_{\Gamma}\ln\left(\frac{q_{\Gamma}}{p_{\Gamma}}\right)
\end{equation}
as the relative entropy or Kullback-Leibler divergence of the path weights~\cite{cover2012elements}.
The argument here, while not formal, is much in the spirit of Sanov's theorem for the empirical density of independent random variables~\cite{touchette2009large}. 

Ultimately, we are interested in odd currents $J_T$, which can be expressed as trajectory averages
\begin{equation}
J_T=\sum_\Gamma {\mathcal F}_\Gamma q_\Gamma,
\end{equation}
whose odd-parity under time-reversal is captured by the symmetry with the time-reversed weight $\tilde{\mathcal F}_\Gamma$:
\begin{equation}\label{symrel}
{\mathcal F}_\Gamma=-\tilde{\mathcal F}_{\tilde \Gamma}.
\end{equation}
Fluctuations in $J_T$ can now readily be obtained from our level 2 large deviation function for independent trajectories through the contraction principle~\cite{touchette2009large}
\begin{equation}
    \mathcal{I}(J_T)=\min_{\{q_{\Gamma}\}} \mathcal{J}(\{q_{\Gamma}\}),
\end{equation}
where the minimization is over distributions $q_{\Gamma}$ that satisfy normalization $ \sum_{\Gamma}q_{\Gamma}=1$ and give the correct flux $\sum_{\Gamma}\mathcal{F}_{\Gamma}q_{\Gamma}=J_T$.
From this large deviation function, many important properties of the current fluctuations can be deduced.
For our purposes, the mean (most likely) current $\langle J_T\rangle$ is the minimum of the large deviation function, which by convention is set to zero: ${\mathcal I}(\langle J_T\rangle)={\mathcal I}'(\langle J_T\rangle)=0$.
Furthermore, the current variance is obtained from the curvature at the minimum~\cite{touchette2009large}
\begin{equation}\label{eq:curvature}
{\mathcal I}''(\langle J_T\rangle) = \frac{1}{{\rm Var}(J_T)}.
\end{equation} 

Exactly the same considerations hold for the large deviations of the reverse process with its empirical density $\tilde{q}_\Gamma$ and  reverse currents ${\tilde J}_T=\sum_\Gamma \tilde{\mathcal F}_\Gamma {\tilde q}_\Gamma$.

\section{Main Result}

\subsection{Fluctuations of hysteretic currents}

Our main inequality is a bound on the joint fluctuations of the hysteretic currents, which we address by considering the joint fluctuations of the forward and reverse processes.

Now the fluctuations in the forward process and the reverse process are independent.
As such, the joint probability for the empirical density $q_\Gamma$ and the empirical density for the reverse process ${\tilde q}_\Gamma$  factorizes ${\mathcal P}_N(\{q_\Gamma\},\{{\tilde q}_{\Gamma'}\})={\mathcal P}_N(\{q_\Gamma\})\tilde{\mathcal P}_N(\{{\tilde q}_{\Gamma'}\})$.
As such, the joint large deviation rate function 
\begin{equation}
-\lim_{N\to\infty}\frac{1}{N}\ln{\mathcal P}_N(\{q_\Gamma\},\{{\tilde q}_{\Gamma'}\})={\mathcal J}(\{q_\Gamma\})+\tilde{\mathcal J}(\{{\tilde q}_{\Gamma'}\})
\end{equation}
is simply the sum of the large deviation functions of the original process ${\mathcal J}$ and the reverse process $\tilde{\mathcal J}$.
With this large deviation function we can extract the fluctuations of the hysteretic currents,
\begin{equation}
J_T+{\tilde J}_T=\sum_\Gamma {\mathcal F}_\Gamma q_\Gamma + \sum_{\Gamma'}\tilde{\mathcal F}_{\Gamma'} {\tilde q}_{\Gamma'},
\end{equation}
from the contraction principle
\begin{equation}\label{eq:hystLDF}
    \mathcal{I}(J_T+{\tilde J}_T)=\min_{\{q_{\Gamma},{\tilde q}_{\Gamma'}\}} \left[\mathcal{J}(\{q_{\Gamma}\})+\tilde{\mathcal J}(\{{\tilde q}_{\Gamma'}\})\right],
\end{equation}
where minimization is constrained by 
\begin{equation}
    \sum_{\Gamma}q_{\Gamma}=\sum_{\Gamma'}\tilde{q}_{\Gamma'}=1,\qquad\sum_\Gamma {\mathcal F}_\Gamma q_\Gamma + \sum_{\Gamma'}\tilde{\mathcal F}_{\Gamma'} {\tilde q}_{\Gamma'}=J_T+{\tilde J}_T.\label{cond}
\end{equation}

\subsection{Uncertainty bound for hysteretic currents}

While the minimization in Eq.~\eqref{eq:hystLDF} can rarely be done analytically, a useful upper bound can be found using the variational ansatz -- which is a slight variation of the ansatz used in \cite{proesmans2017discrete} to obtain Eq.~\eqref{tur2} -- 
\begin{eqnarray}
    q_{\Gamma}^{\left(J_T+{\tilde J}_T\right)}&=&p_{\Gamma}+\frac{J_T+{\tilde J}_T-\left\langle J_T+{\tilde J}_T\right\rangle}{\left\langle J_T+{\tilde J}_T\right\rangle}\left(p_{\Gamma}-\frac{p_{\Gamma}{\tilde p}_{\tilde{\Gamma}}}{\mathcal{N}\left(p_{\Gamma}+{\tilde p}_{\tilde{\Gamma}}\right)}\right)\nonumber\\
    {\tilde q}^{\left(J_T+{\tilde J}_T\right)}_\Gamma &=&{\tilde p}_{\Gamma}+\frac{J_T+{\tilde J}_T-\left\langle J_T+{\tilde J}_T\right\rangle}{\left\langle J_T+{\tilde J}_T\right\rangle}\left({\tilde p}_{\Gamma}-\frac{p_{\tilde{\Gamma}}{\tilde p}_{\Gamma}}{\mathcal{N}\left(p_{\tilde{\Gamma}}+{\tilde p}_{\Gamma}\right)}\right),\label{qcho}
\end{eqnarray}
with
\begin{equation}
    \mathcal{N}=\sum_{\Gamma}\frac{p_{\Gamma}{\tilde p}_{\tilde{\Gamma}}}{p_{\Gamma}+{\tilde p}_{\tilde{\Gamma}}},\label{Ndef}
\end{equation}
{ where we emphasize that ${\tilde p}_{\tilde\Gamma}$ is the probability in the reverse process to observe $\tilde\Gamma$, the time reverse of the trajectory $\Gamma$.}
{ The rationale for this ansatz can be seen by recognizing it as an expansion of $q_\Gamma$ around its most likely value $p_\Gamma$ to first order in the deviation in the hysteric current around its typical value, while maintaining the constraints.
While many ansatzs could satisfy this criteria, this one minimizes the resulting bound on the variance. }

Using Eq.~(\ref{symrel}), it is clear that the conditions in Eq.~(\ref{cond}) are valid, and therefore,
\begin{eqnarray}
\mathcal{I}(J_T+{\tilde J}_T)&\leq& \mathcal{J}(\{q^{\left(J_T+{\tilde J}_T\right)}_{\Gamma}\})+\tilde{\mathcal{J}}(\{{\tilde q}^{\left(J_T+{\tilde J}_T\right)}_{\Gamma}\}).
\end{eqnarray}
This translates into a bound on variance through Eq.~\eqref{eq:curvature} after noting that $\mathcal{J}(\{q^{\left(\langle J_T+{\tilde J}_T\rangle\right)}_{\Gamma}\})=\mathcal{J}'(\{q^{\left(\langle J_T+{\tilde J}_T\rangle\right)}_{\Gamma}\})=0$ (and analogous relations for $\tilde{\mathcal{J}}$),
\begin{eqnarray}
   \mathcal{I}''(\langle J_T+{\tilde J}_T\rangle)=\frac{1}{\textrm{Var}(J_T+{\tilde J}_T)}
    \leq \frac{1}{\left\langle J_T+{\tilde J}_T\right\rangle^2}\left(\frac{1}{\mathcal{N}}-2\right).\label{varbound}
\end{eqnarray}

To arrive at the final result, we now need to bound $\mathcal{N}$. 
Following closely \cite{proesmans2017discrete}, we introduce the pair of trajectory averages with an eye towards entropy flow (cf.~Eq.~\eqref{eq:Ent}),
\begin{equation}\label{eq:R}
{\mathcal R}+\tilde{\mathcal R}  = \sum_\Gamma p_\Gamma \ln\frac{p_\Gamma}{{\tilde p}_{\tilde \Gamma}}+\sum_{\Gamma'} {\tilde p}_{\Gamma'} \ln\frac{{\tilde p}_{\Gamma'}}{p_{\tilde \Gamma'}}.
\end{equation}
Now from the definition of ${\mathcal N}$ (Eq.~(\ref{Ndef})) and an application of Jensen's inequality we have
\begin{eqnarray}
\ln\mathcal{N}+\frac{{\mathcal R}+\tilde{\mathcal R}}{2}&=&\ln\left(\sum_{\Gamma}\frac{p_{\Gamma}{\tilde p}_{\tilde{\Gamma}}}{p_{\Gamma}+{\tilde p}_{\tilde{\Gamma}}}\right)+\sum_{\Gamma}\frac{p_{\Gamma}-{\tilde p}_{\tilde{\Gamma}}}{2}\ln\left(\frac{p_{\Gamma}}{{\tilde p}_{\Gamma}}\right)\nonumber\\
&\geq&\sum_{\Gamma}\frac{p_{\Gamma}+{\tilde p}_{\tilde{\Gamma}}}{2}\ln\left(\frac{2p_{\Gamma}{\tilde p}_{\tilde{\Gamma}}}{(p_{\Gamma}+{\tilde p}_{\tilde{\Gamma}})^2}\right)\nonumber+\sum_{\Gamma}\frac{p_{\Gamma}-{\tilde p}_{\tilde{\Gamma}}}{2}\ln\left(\frac{p_{\Gamma}}{{\tilde p}_{\tilde{\Gamma}}}\right)\nonumber\\
&=&\sum_{\Gamma} {\tilde p}_{\tilde\Gamma}\left(\frac{1+u_{\Gamma}}{2}\ln\frac{2u_{\Gamma}}{(1+u_{\Gamma})^2}+\frac{u_{\Gamma}-1}{2}\ln u_{\Gamma}\right),\label{ineq}
\end{eqnarray}
with $u_{\Gamma}=p_{\Gamma}/{\tilde p}_{\tilde{\Gamma}}$. Using the one-dimensional inequality\cite{proesmans2017discrete,barato2018unifying}
\begin{equation}
    \frac{1+u}{2}\ln\frac{2u}{(1+u)^2}+\frac{u-1}{2}\ln u\geq(1-\ln 2)\frac{1+u}{2}-\frac{2u}{u+1},
\end{equation}
which can be verified upon graphical inspection, one can do the summation in the last line of Eq.~(\ref{ineq}) explicitly, leading to
\begin{equation}
\ln\mathcal{N}+\frac{{\mathcal R}+\tilde{\mathcal R}}{2}\geq 1-\ln 2-2\mathcal{N}\geq\ln(1-\mathcal{N}),
\end{equation}
where the last inequality follows from the concavity of the logarithmic function.  It now readily follows that

\begin{equation}
    \frac{1}{\mathcal{N}}\leq e^{\frac{{\mathcal R}+\tilde{\mathcal R}}{2}}+1.
\end{equation}
Plugging this back in into Eq.~(\ref{varbound}), immediately leads to the final result in Eq.~\eqref{eq:newTUR},
after noting that since the two processes are independent the joint variance is the sum of the individual variances.

\section{Applications}

The path averages ${\mathcal R}+\tilde{\mathcal R}$ in Eq.~\eqref{eq:R} are intimately related to the entropy flow in Eq.~\eqref{eq:Ent}.
Indeed, we can peel off the entropy flow contribution as follows.
For ${\mathcal R}$ (and analogously for $\tilde{\mathcal R}$), we have
\begin{equation}\label{eq:R2}
{\mathcal R} = \Delta S_{\rm env}/k_{\rm B}+{\mathcal B}
\end{equation}
where we have singled out the boundary contribution
\begin{equation}\label{eq:B}
{\mathcal B}=\left\langle\ln\frac{\rho_0({\bf x}(0))}{\tilde\rho_0(\boldsymbol\epsilon {\bf x}(T))}\right\rangle_{\boldsymbol\lambda (t)}=\sum_{\bf x}\rho_0({\bf x})\ln\rho_0({\bf x})-\sum_{\bf x}\rho_T({\bf x})\ln \tilde\rho_0(\boldsymbol\epsilon {\bf x}),
\end{equation}
with $\rho_T$ the final distribution of the forward process.
Roughly speaking, this boundary contribution ${\mathcal B}$ quantifies the time-reversal asymmetry between the initial and terminal distributions of the two processes.
Though to provide ${\mathcal B}$ with a clear thermodynamic interpretation, we need to specify these distributions.
In this section, we explore some possible choices.

\subsection{Isothermal work fluctuations}

Let us specialize to systems coupled to a single thermal reservoir at inverse temperature $\beta$.
In the absence of driving, we assume the system relaxes to a thermal equilibrium state given by the Botlzmann distribution $\rho^{\rm eq}({\bf x};\boldsymbol\lambda)= e^{\beta(F(\boldsymbol\lambda) -E({\bf x};\boldsymbol\lambda))}$ with time-reversal invariant energy function $E({\bf x};\boldsymbol\lambda)=E(\boldsymbol\epsilon{\bf x};\boldsymbol\epsilon \boldsymbol\lambda)$ and free energy $F(\boldsymbol\lambda$).

An interesting choice for initial conditions in this situation is to choose the initial distribution of the forward process to be the initial equilibrium distribution $\rho_0({\bf x})=\rho^{\rm eq}({\bf x};\boldsymbol\lambda(0))$ and for the reverse process the final equilibrium distribution $\tilde\rho_0({\bf x})=\rho^{\rm eq}({\bf x};\boldsymbol\lambda(T))$~\cite{Jarzynski2011}.
In this case,
\begin{equation}
{\mathcal R}+\tilde{\mathcal R}=\beta\langle W_T + \tilde W_T\rangle,
\end{equation}
is simply the sum of the work into the system in the forward and reverse processes, \emph{i.e.}, the work hysteresis~\cite{feng2008length,crooks2011measures}.

A particular interesting current is then the work hysteresis itself, $W_T+{\tilde W}_T$, leading to the thermodynamic bound on work fluctuations
\begin{equation}\label{eq:work}
    \frac{\left\langle W_T+ \tilde{W}_T\right\rangle^2}{\textrm{Var}(W_T)+\textrm{Var}(\tilde{W}_T)}\leq \exp\left(\frac{\beta}{2}\langle W_T+ \tilde{W}_T\rangle\right)-1.
\end{equation}
It should be noted that this bound on work fluctuations is independent of the work fluctuation relations~\cite{Jarzynski2011}, offering an orthogonal constraint to the structure of work distributions.

\subsection{Steady-state fluctuations within linear response \label{lintur}}

Another special case where progress can be made is systems at near-equilibrium steady states.
Let us assume that when the parameters are time-independent $\boldsymbol\lambda$, the system relaxes to a parameter-dependent steady-state distribution $\pi({\bf x};\boldsymbol\lambda)$.
By choosing the initial distribution of the forward process as the steady-state distribution $\rho_0({\bf x})=\pi({\bf x};\boldsymbol\lambda)$ and initiating the reverse process in the parameter-reversed steady-state distribution $\tilde\rho_0({\bf x})=\pi({\bf x};\boldsymbol\epsilon{\boldsymbol\lambda})$ we find that the boundary contributions sum to
\begin{align}
{\mathcal B}+\tilde{\mathcal B}=\sum_{\bf x}\pi({\bf x};\boldsymbol\lambda)\ln \frac{\pi({\bf x};\boldsymbol\lambda)}{\pi(\boldsymbol\epsilon{\bf x};\boldsymbol\epsilon\boldsymbol\lambda)}
+\sum_{\bf x}\pi({\bf x};\boldsymbol\epsilon\boldsymbol\lambda)\ln \frac{\pi({\bf x};\boldsymbol\epsilon \boldsymbol\lambda)}{\pi(\boldsymbol\epsilon{\bf x};\boldsymbol\lambda)},
\end{align}
which is a symmetrized relative entropy (Jensen-Shannon divergence \cite{cover2012elements}) of the steady-state distribution with its time reverse.
Thus, when the steady-state distribution is time-reversal symmetric these boundary terms drop out.
This term has previously been identified in \cite{spinney2012nonequilibrium,Granger2015} as an unavoidable component in the analysis of the thermodynamics of systems with broken time-reversal symmetry.

{ It is now very natural, in light of the original thermodynamic uncertainty relation in Eq.~\eqref{tur1}, to consider how our bound constrains the long time limit of the current fluctuations: $\langle J_T\rangle= T \langle j\rangle $ and ${\rm Var}(J_T)\to T\,  \textrm{Var}(j)$.
Unfortunately, our result turns out not to be  informative in this limit: for long times the entropy flow grows linearly with time $\Delta S_{\rm env}=T\sigma$ causing the our upper bound to diverge.
Thus, despite the close formal similarity with previous results, direct comparison is challenging.
Interesting progress can be made however for systems near equilibrium within linear response.}

Within linear response all currents and conjugate thermodynamic forces are small, but their fluctuations need not be.
To formalize this notion, it is helpful to introduce a small parameter $\eta$ that `measures' the distance from equilibrium; roughly, the strength of the thermodynamic forces holding the system slightly out of equilibrium.
As the entropy flow is a bilinear function of currents and forces, it too is small but behaves as $\sim\eta^2$.
Similarly, one can show using standard perturbation theory that if the steady-state is near equilibrium as assumed, $\pi({\bf x};\boldsymbol\lambda)=\rho^{\rm eq}({\bf x},\boldsymbol\lambda)+\eta\ \delta\pi({\bf x};\boldsymbol\lambda)$, then the boundary terms as relative entropies also behave as $\sim\eta^2$.
Thus, we can make a consistent expansion of our uncertainty relation in Eq.~\eqref{eq:newTUR} utilizing our identifications in Eqs.~\eqref{eq:R2} and \eqref{eq:B} resulting in the near-equilibrium bound
\begin{equation}
\frac{\langle J_T+{\tilde J}_T\rangle^2}{{\rm Var}(J_T)+{\rm Var}({\tilde J}_T)}\le \frac{\Delta S_{\rm env}+\Delta {\tilde S}_{\rm env}}{2k_{\rm B}}+\frac{\delta {\mathcal B}+\delta \tilde{\mathcal B}}{2},
\end{equation}
with $\delta{\mathcal B}$ and $\delta{\tilde{\mathcal B}}$ the lowest-order linear-response corrections to the boundary terms, which are zero for time-reversal symmetric steady-states.

Interestingly, the steady-state version of this linear response inequality is reminiscent of the original formulation of the uncertainty relation.
Indeed, { returning to the} steady-state current $\langle j\rangle =\lim_{T\to\infty}\langle J_T\rangle/T$, variance $\textrm{Var}(j)=\lim_{T\to\infty}{\rm Var}(J_T)/T$, and entropy flow which becomes equal to the entropy production rate in the steady state $\sigma =\lim_{T\to\infty}\Delta S_{\rm env}/T$, we have 
\begin{equation}\label{eq:TURlr}
    \frac{\left\langle j+\tilde{j}\right\rangle^2}{\textrm{Var}(j)+\textrm{Var}\left({\tilde j}\right)}\leq \frac{\sigma+\tilde\sigma}{2k_B},
\end{equation}
after noting that the boundary terms are not extensive in time. In \ref{app1}, we provide an alternative viewpoint on this prediction by deriving it entirely within the framework of linear response theory.

\section{Illustrations \label{ex}}
In this section, we provide two illustrations that highlight the utility of the two applications discussed in the previous section.
 The first example is a model for the adiabatic expansion/compression of a dilute, weakly-interacting gas.
 The second is a steady-state multi-terminal conductor in an external magnetic field within linear response. 

\subsection{Isothermal work protocol}
\begin{figure}[htb]
\begin{center}
\includegraphics[scale=0.4]{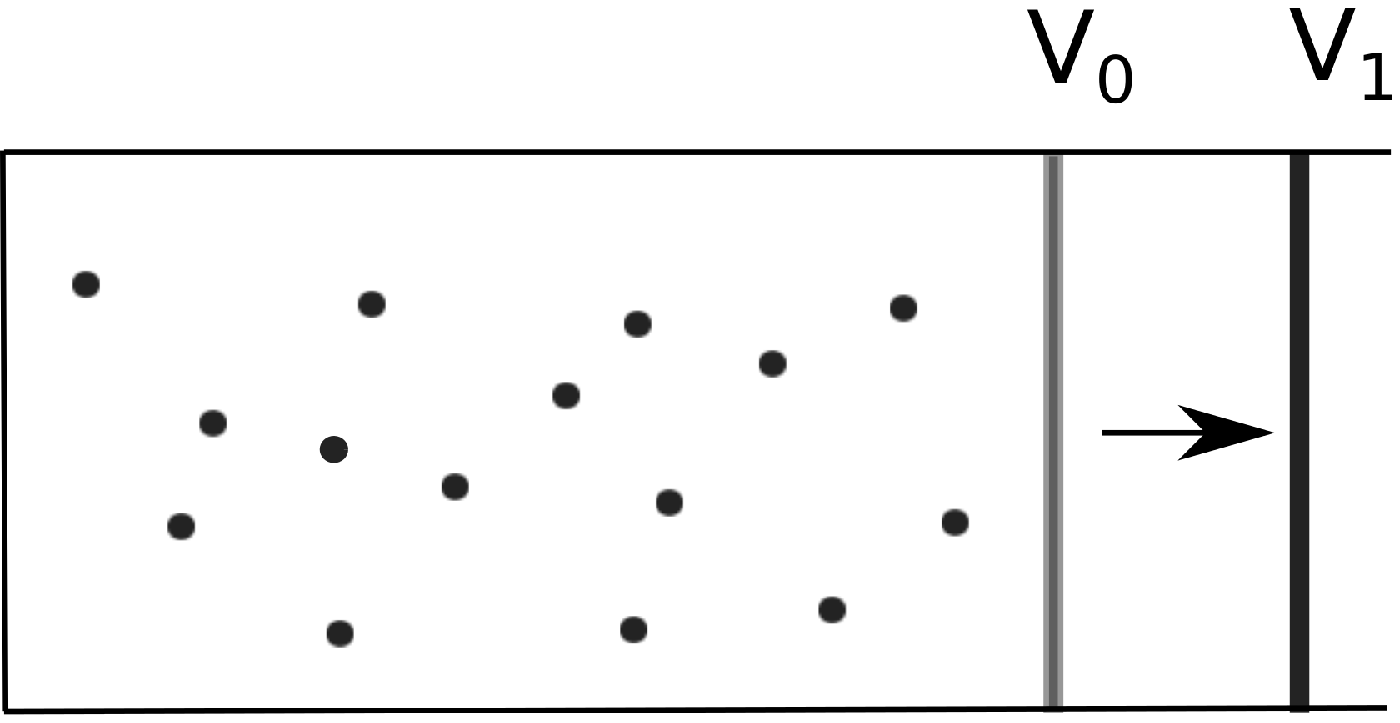}
\qquad
\includegraphics[scale=0.4]{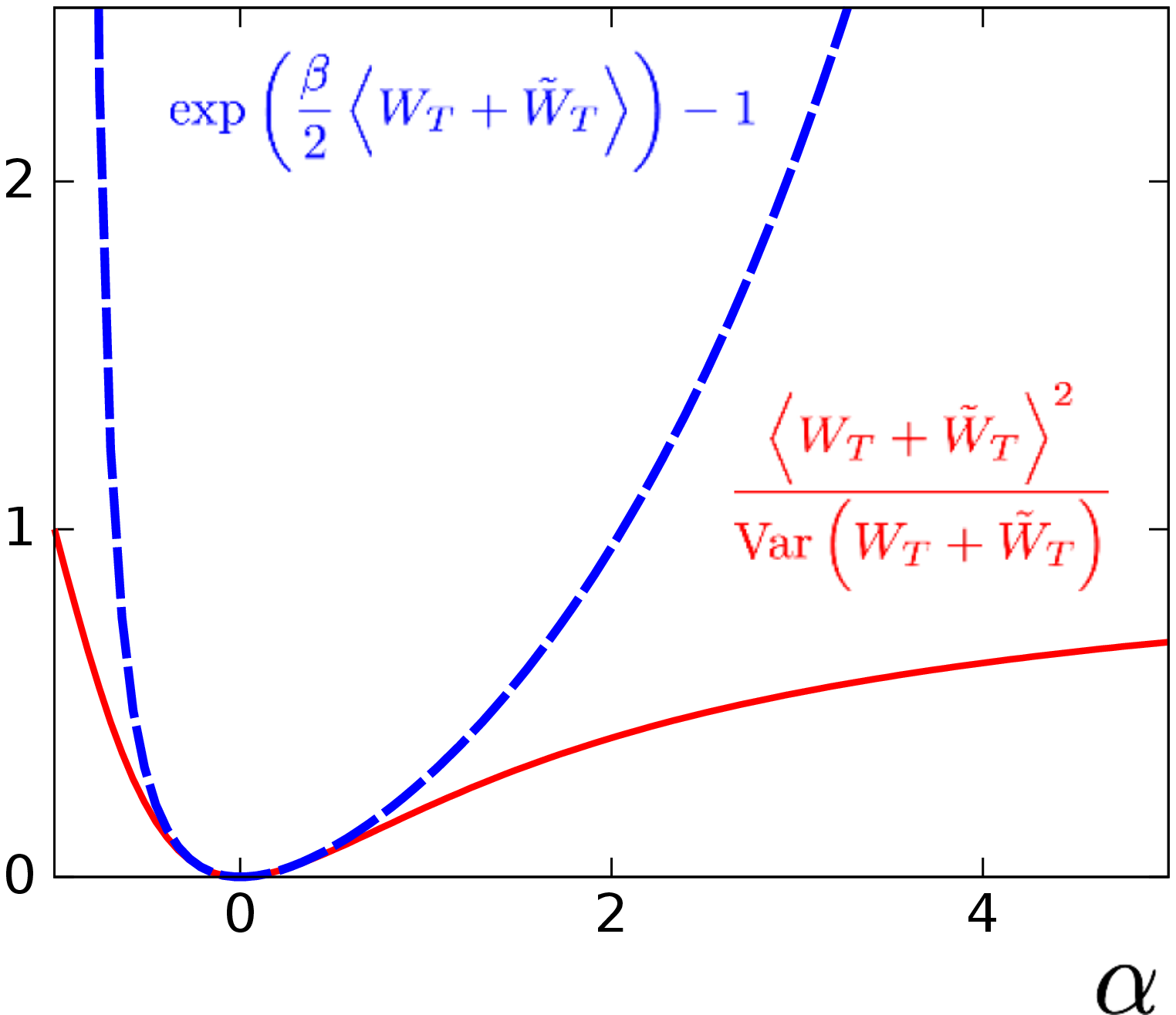}
\end{center}
\caption{Left panel: Depiction of a dilute gas confined to a box with a movable piston used to expand the volume from $V_0$ to $V_1$. Right panel: Verification of the work uncertainty relation for $d=2$, $N=1$ as a function of the work-distribution parameter $\alpha = (V_0/V_1)^{2/d}-1$.}
\label{congas}
\end{figure}

Our first illustration is a solvable model developed by Crooks and Jarzynski~\cite{crooks2007work}.
The system of interest is a dilute, weakly-interacting gas of $N$ particles confined to a box in $d$ dimensions with one wall actuated by a movable piston that allows one to do work by varying the volume, Fig.~\ref{congas}.
The gas is initialized in thermal equilibrium at inverse temperature $\beta$ and volume $V_0$ by coupling the gas to a large thermal reservoir.
The reservoir is then disconnected  and the piston is slowly, adiabatically compressed (or expanded) changing the volume from $V_0\to V_1$.

In this model, the full probability distribution of work can be determined analytically~\cite{crooks2007work}, and is given by the Gamma distribution parameterized by $\alpha = (V_0/V_1)^{2/d}-1$ as
\begin{equation}
P(W)=\frac{\beta}{|\alpha|\Gamma(k)}\left(\frac{\beta W}{\alpha}\right)^{k-1}e^{-\beta W/\alpha}\theta(\alpha W),
\end{equation}
with shape parameter $k=d N/2$, where $\Gamma(k)$ is the Gamma function and $\theta(y)$ is the Heaviside step function.
For example, a compression has $V_0>V_1$ implying $\alpha>0$ and the work distribution is supported only on the positive work values.

We will now consider the forward process to be compression from $V_0\to V_1$ with associated parameter $\alpha=(V_0/V_1)^{2/d}-1$, and the reverse process is expansion from $V_1\to V_0$ with $\tilde\alpha = (V_1/V_0)^{2/d}-1$.
Reading off the mean $\langle W_T\rangle = k\alpha/\beta$ and variance ${\rm Var}(W_T)=k\alpha^2/\beta^2$ from the Gamma distribution, one can check that the work uncertainty relation Eq.~\eqref{eq:work},
\begin{equation}
 \frac{dN}{2}\frac{(\alpha+\tilde\alpha)^2}{\alpha^2+\tilde\alpha^2}\le \exp\left(\frac{dN}{4}(\alpha+\tilde\alpha)\right)-1,
\end{equation}
is valid as $\tilde{\alpha}=(1+\alpha)^{-1}-1$. This result is illustrated in Fig.~\ref{congas}.

\subsection{Multi-terminal conductor}

\begin{figure}[htb]
\begin{center}
\includegraphics[scale=0.6]{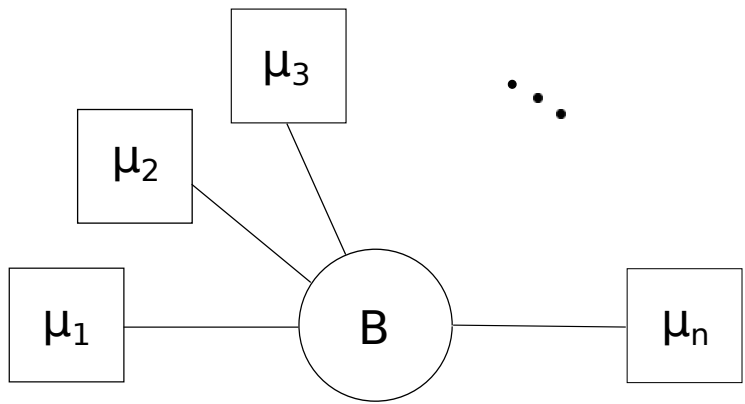}
\end{center}
\caption{Illustration of a multi-terminal conductor: A central scattering region (circle) in a magnetic field $B$ exchanges particles with $n$ particle reservoirs with chemical potentials $\mu_1,\dots,\mu_n$ and equal inverse temperature $\beta$.}
\label{setupmtc}
\end{figure}

Consider a collection of $n$ particle reservoirs, connected via a central scattering region in a magnetic field $B$, where each reservoir $\alpha$ has a constant inverse temperature $\beta$ and chemical potential $\mu_\alpha$, $\alpha=1,\dots,n$, c.f.~Fig.~\ref{setupmtc}. This class of systems was studied previously in the context of thermodynamic uncertainty relations in \cite{PhysRevLett.121.130601}, where it was shown that the original uncertainty relation Eq.~\eqref{tur1} does not hold due to the presence of the magnetic field. 

We are interested in the particle flux to each reservoir $\alpha$. This can be written in terms of the transmission matrix $\mathcal{T}^{\alpha\gamma}_{\mathbf{B}}(E)$, which gives the rate at which a particle starting at reservoir $\alpha$ with energy $E$ ends up in reservoir $\gamma$. The flux is then given by the multi-terminal Landauer formula \cite{sivan1986multichannel,butcher1990thermal,brandner2013multi},
\begin{equation}
    \left\langle j_\alpha\right\rangle=\frac{1}{h}\int^{\infty}_0dE\,\sum_\gamma \mathcal{T}^{\alpha\gamma}_{\mathbf{B}}(E)(u^{\alpha}(E)-u^{\gamma}(E)),
\end{equation}
where $h$ is Planck's constant and
\begin{equation}
    u^{\alpha}(E)=\exp\left[-\beta(E-\mu_\alpha)\right].
\end{equation}
One can show that the variance associated with the particle flux is given by \cite{brandner2018thermodynamic}
\begin{equation}
    \textrm{Var}(j_\alpha)=\frac{1}{h}\int^{\infty}_0dE\, \sum_{\gamma\neq \alpha}\mathcal{T}^{\alpha\gamma}_{\mathbf{B}}(E)(u^{\alpha}(E)+u^{\gamma}(E)),
\end{equation}
and the average entropy production rate is
\begin{equation}
    \sigma=\sum_\alpha {F}_\alpha j_\alpha,
\end{equation}
with
\begin{equation}
    {F}_\alpha=\beta(\mu_\alpha-\mu_0),
\end{equation}
$\mu_0$ being an arbitrary, constant chemical potential. 

In the time-reversed situation, where the magnetic field reverses its sign, the above expressions stay the same apart from the transmission matrix $\tilde{\mathcal{T}}^{\alpha\gamma}_\mathbf{B}(E)={\mathcal{T}}^{\alpha\gamma}_\mathbf{-B}(E)$. 
The transmission matrix does however satisfy an important Onsager-Casimir-like symmetry relation \cite{brandner2013multi},
\begin{equation}
    \mathcal{T}^{\alpha\gamma}_\mathbf{-B}(E)=\mathcal{T}^{\gamma\alpha}_\mathbf{B}(E).\label{Tsym}
\end{equation}

From these basic expressions, formulas for the hysteretic fluctuations follow readily.
They are conveniently expressed through the quantities~\cite{PhysRevLett.121.130601}
\begin{equation}
    \mathcal{V}^{\alpha\gamma}_\mathbf{B}=\frac{1}{h}\int^{\infty}_0dE\, \left(\mathcal{T}^{\alpha\gamma}_\mathbf{B}(E)+\tilde{\mathcal{T}}^{\alpha\gamma}_\mathbf{B}(E)\right)u^{\gamma}(E),\qquad \mathcal{D}_{\alpha\gamma}={{F}_\alpha-{F}_\gamma}
\end{equation}
as
\begin{align}\label{eq:cond1}
\left\langle j_\alpha+\tilde{j}_\alpha\right\rangle&=\sum_\gamma\mathcal{V}_\mathbf{B}^{\alpha\gamma}\left(e^{\mathcal{D}_{\alpha\gamma}}-1\right)\\
\textrm{Var}(j_\alpha)+\textrm{Var}(\tilde{j}_\alpha)&=\sum_{\gamma\neq \alpha}\mathcal{V}^{\alpha\gamma}_\mathbf{B}\left(e^{\mathcal{D}_{\alpha\gamma}}+1\right)\\
\sigma+\tilde{\sigma}&=\frac{k_B}{2}\sum_{\alpha,\gamma}\mathcal{V}^{\alpha\gamma}_\mathbf{B}\mathcal{D}_{\alpha\gamma}\left(e^{\mathcal{D}_{\alpha\gamma}}-1\right).
\end{align}

We can now verify the linear response uncertainty relation in Eq.~\eqref{eq:TURlr} by demonstrating that $X={(\sigma+\tilde{\sigma})(\textrm{Var}(j_\alpha)+\textrm{Var}(\tilde{j}_\alpha))}/({2k_B})-\left\langle j_\alpha+\tilde{j}_\alpha\right\rangle^2$ is positive.
To this end, we first note that $\mathcal{V}_{\mathbf{B}}^{\alpha\gamma}\left(e^{\mathcal{D}_{\alpha\gamma}}-1\right)=-\mathcal{V}_{\mathbf{B}}^{\gamma\alpha}\left(e^{\mathcal{D}_{\gamma\alpha}}-1\right)$, and therefore
\begin{equation}\label{eq:cond4}
    \sigma+\tilde{\sigma}\geq\frac{k_B}{h}\sum_{\gamma}\mathcal{V}^{\alpha\gamma}_\mathbf{B}\mathcal{D}_{\alpha\gamma}\left(e^{\mathcal{D}_{\alpha\gamma}}-1\right).
\end{equation}
Now combining Eqs.~\eqref{eq:cond1}-\eqref{eq:cond4}, we find
\begin{align}
   X &\geq \frac{1}{h}\sum_{\gamma,\gamma'}\mathcal{V}_\mathcal{B}^{\alpha\gamma}\mathcal{V}_\mathcal{B}^{\alpha\gamma'}\left(\frac{\mathcal{D}_{\alpha\gamma}}{2}\left(e^{\mathcal{D}_{\alpha\gamma}}-1\right)\left(e^{\mathcal{D}_{\alpha\gamma'}}+1\right)-\left(e^{\mathcal{D}_{\alpha\gamma}}-1\right)\left(e^{\mathcal{D}_{\alpha\gamma'}}-1\right)\right)\nonumber\\
    &\geq\frac{2}{h}\sum_{\gamma,\gamma'}\mathcal{V}_\mathcal{B}^{\alpha\gamma}\mathcal{V}_\mathcal{B}^{\alpha\gamma'}\frac{\left(e^{\mathcal{D}_{\alpha\gamma}}-e^{\mathcal{D}_{\alpha\gamma'}}\right)^2}{(e^{\mathcal{D}_{\alpha\gamma}}+1)(e^{\mathcal{D}_{\alpha\gamma'}}+1)}\nonumber\\
    &\geq 0,
\end{align}
where we used the inequality $y(\exp(y)-1)\geq 2(\exp(y)-1)^2/(\exp(y)+1)$. 

\section{Conclusion \label{con}}
In this paper, we derived a finite-time thermodynamic uncertainty relation for systems with broken time-reversal symmetry by looking at hysteretic currents -- joint current fluctuations in the time-forward and time-reversed processes.
Our result implies that for nonequilibrium steady-states within linear response, the original thermodynamic uncertainty relation holds as long as we account for the time-reversal asymmetry of the steady state.
When this hysteretic uncertainty relation is specialized to driven isothermal work processes, it implies a nontrivial relationship between the fluctuations of the work in the forward and reverse process.
In each case, the quantities have clear physical and thermodynamic interpretations.
We have also verified that the hysteretic uncertainty relation holds for two examples that strictly do not fall within our setup: a ballistic multi-terminal conductor as well as for slow, adiabatic compression of a gas.
 This raises an interesting question as to the generality of our extension, a question which might be answered by doing a fully Hamiltonian derivation of the thermodynamic uncertainty relation.

 We have also seen that direct comparison with earlier results for steady-state currents arbitrarily far from equilibrium is challenged by the structure of our bound.
 A similar difficulty arrises when comparing with uncertainty bounds derived for system driven by time-periodic protocols, where the issue of taking the limit of number of periods to infinity leads to an uninformative inequality.
Indeed, in \cite{1367-2630-20-10-103023,barato2018unifying,koyuk2018generalization}, relations similar to Eqs.~(\ref{tur1}) and (\ref{tur2}) were derived for systems with time-asymmetric driving, but focused on this long-time limit.
On the other hand, our finite-time results only depends on the entropy production, while previous results mixed entropy production with various other quantities.  Some use measures of the time-reversal symmetry breaking in the kinetics.
Others introduce entropy-production-like quantities but utilizing time-averaged current and forces.
 It would be interesting to see how our bound compares to these results from the literature more precisely.

Beyond the cases considered here one can think of several other classes of systems where the hysteretic uncertainty relation can give new physical insights. One could for example think about systems which are prepared in echo states \cite{becker2015echo}. Another interesting extension would be to quantum systems, where one might overcome some of the observed deficiencies in the original uncertainty relation \cite{brandner2018thermodynamic,PhysRevB.98.085425}.
It would also be interesting to see if one can use the hysteretic uncertainty relation to constrain the performance of heat engines with broken time-reversal symmetry, in a similar way as was done for time-symmetric steady-state driving \cite{pietzonka2018universal}.

\section*{Acknowledgment}
The authors acknowledge the Max Planck Institute for the Physics of Complex Systems which hosted the International Workshop ``Stochastic Thermodynamics: Experiment and Theory'' where this work was initiated as well as Hugo Touchette for useful remarks.
KP is a postdoctoral fellow of the Research Foundation-Flanders (FWO).
JMH is supported by the Gordon and Betty Moore Foundation as a Physics of Living Systems Fellow through Grant No. GBMF4513.

\appendix
\section{Thermodynamic uncertainty relation near equilibrium \label{app1}}
In this appendix, we give an alternative derivation of the hysteretic thermodynamic uncertainty relation in the linear response regime, presented in section \ref{lintur}, using linear response theory \cite{de2013non}.  Previous derivations of the time-symmetric uncertainty relation within linear response~\cite{barato2015thermodynamic,PhysRevLett.121.130601} relied on the Onsager matrix being positive-definite.  Similar to these approahes, we will see by using hysteretic currents, that we can again exploit the fact that the symmetric part of the Onsager-Casimir matrix is positive-definite. (In general, the Onsager-Casimir matrix itself is not positive-definite.) 

For notational simplicity, we will focus on a system with two independent thermodynamic fluxes $J_1$ and $J_2$ with thermodynamic forces ${F}_1$ and ${F}_2$ conjugated through the entropy production rate $\sigma=k_B (J_1{F}_1+J_2{F}_2)$. The following can easily be extended to systems with more thermodynamic fluxes. In the linear response regime, the currents can be expanded linearly in terms of the forces \cite{de2013non}
\begin{equation}
   J_1=L_{11}{F}_1+L_{12}{F}_2,\qquad  J_2=L_{21}{F}_1+L_{22}{F}_2,\label{Jdef}
  \end{equation}
where the expansion parameters $L_{ij}$ are known as the Onsager coefficients. 
 In terms of these coefficients, the entropy production can be written as the quadratic form
\begin{equation}
    \sigma=k_B\sum_{i,j=1}^2{F}_iL_{ij}{F}_j.
\end{equation}
Notably, if we introduce the symmetric and antisymmetric Onsager coeffecients
\begin{equation}
    L_{ij}^{(s)}=\frac{L_{ij}+L_{ji}}{2},\qquad L_{ij}^{(a)}=\frac{L_{ij}-L_{ji}}{2},
\end{equation}
then the entropy production only depends on the symmetrized coeffecients, $\sigma=k_B\sum_{i,j=1}^2{F}_iL^{(s)}_{ij}{F}_j$.
The second law then guarantees the positivity of the entropy production independent of the choice of ${\mathcal F}_i$, which implies the useful condition on the symmetric Onsager coefficients
\begin{equation}
    L^{(s)}_{11}L^{(s)}_{22}\geq \left.L_{12}^{(s)}\right.^2.
\end{equation}

Analogously, we can write for the time-reversed dynamics (assuming the thermodynamic forces to be time-reversal symmetric),
\begin{equation}
    \tilde{J}_1=\tilde{L}_{11}{F}_1+\tilde{L}_{12}{F}_2,\qquad  \tilde{J}_2=\tilde{L}_{21}{F}_1+\tilde{L}_{22}{F}_2
\end{equation}
where the time-reversed Onsager matrix can be related to the time-forward Onsager matrix via Onsager-Casimir symmetry \cite{de2013non},
\begin{equation}
    \tilde{L}_{ij}=L_{ji}.
\end{equation}
The Onsager-Casimir symmetry further determines the symmetric and anti-symmetric Onsager coefficients associated with the time-reversed driving
\begin{equation}
    \tilde{L}_{ij}^{(s)}=L_{ij}^{(s)},\qquad \tilde{L}_{ij}^{(a)}=-L_{ij}^{(a)}.
\end{equation}
As such, within linear response the entropy production of time-forward and time-reversed dynamics are equal:
\begin{equation}
\tilde\sigma\equiv k_B\sum_{i,j=1}^2{\tilde J}_i{F}_i=k_B\sum_{i,j=1}^2{F}_i\tilde{L}^{(s)}_{ij}{F}_j=\sigma.
\end{equation}

The Onsager-Casimir symmetry significantly simplifies the quantities appearing in the hysteretic uncertainty relation.
Indeed, we have the hysteretic entropy production rate
\begin{equation}
    \sigma+\tilde{\sigma}=2k_B\sum_{i,j}{F}_iL_{ij}^{(s)}{F}_j.
\end{equation}
From the definition of the thermodynamic fluxes, Eq.~(\ref{Jdef}), and the above constrains, we can also write
\begin{equation}
    \left\langle J_{1}+\tilde{J}_1\right\rangle=2 \left(L^{(s)}_{11}{F}_1+L^{(s)}_{12}{F}_2\right),
\end{equation}
and an analogous relation for $J_{2}$. Meanwhile, the variance of the fluxes can be written in terms of the Onsager coefficients via the fluctuation-dissipation relation,
\begin{equation}
    \textrm{Var}(J_1)=2L^{(s)}_{11},
\end{equation}
and therefore,
\begin{equation}
    \textrm{Var}(J_1+\tilde{J}_1)=4L_{11}^{(s)}.
\end{equation}

Combining these results leads to
\begin{eqnarray}
\frac{(\sigma+\tilde{\sigma})\textrm{Var}(J_1+\tilde{J}_1)}{2k_B}-\left\langle J_1+\tilde{J}_1\right\rangle^2&=& 4L^{(s)}_{11}\sum_{j,k}{F}_jL^{(s)}_{jk}{F}_k\nonumber\\&&-4\sum_{k,l}{F}_kL_{1k}^{(s)}L_{1j}^{(s)}{F}_j\nonumber\\&=&4{F}_2^2\left(L_{11}^{(s)}L_{22}^{(s)}-\left.L_{12}^{(s)}\right.^2\right)\nonumber\\
&\geq& 0,
\end{eqnarray}
which verifies Eq.~(\ref{eq:TURlr}).

\section*{Bibliography}
\providecommand{\newblock}{}

\end{document}